\documentclass[12pt,english,twocolumn]{emulateapj}
\usepackage[latin9]{inputenc}                        
\usepackage{amsmath}
\makeatletter
\usepackage{graphicx}
\usepackage{epsfig}\usepackage{color}\usepackage{txfonts}\bibpunct{(}{)}{;}{a}{}{,} 
\usepackage{url}

\usepackage{hyperref}
\shorttitle{Evolution of second generation stellar disks in clusters}\shortauthors{Mastrobuono-Battisti \& Perets}

\makeatother

\begin{document}

\title{Second generation stellar disks in Dense Star Clusters and cluster ellipticities}

\author{Alessandra Mastrobuono-Battisti\altaffilmark{1, 2} \& Hagai B. Perets\altaffilmark{1}}

\altaffiltext{1}{Physics Department, Technion - Israel Institute of Technology, Haifa,
Israel 32000 }
\altaffiltext{2}{Max-Planck Instiut f\"ur Astronomie, K\"onigstuhl 17, 69117 Heidelberg, Germany}
\begin{abstract}
Globular Clusters (GCs) and Nuclear Star Clusters (NSCs) are typically composed by 
several stellar populations, { characterized  by different chemical compositions. Different populations show different ages in NSCs but not necessarily in GCs.}
The youngest populations in NSCs appear to reside in disk-like structures, as observed in our Galaxy and in M31. Gas infall followed 
by formation of second generation (SG) stars in GCs may similarly form disk-like structures in the clusters nuclei. Here we explore 
this possibility and follow the long term evolution of stellar disks embedded in GCs, and study their {  effects} on the evolution of 
the clusters. We study disks with different masses by means of detailed N-body simulations and explore their morphological and 
kinematic signatures on the GC structures. 
We find that as a second generation disk relaxes, the old, first generation, stellar population flattens and becomes more radially 
anisotropic, making the GC structure become more elliptical. The second generation stellar population is characterized by a lower 
velocity dispersion, and a higher rotational velocity, compared with the primordial older population. 
The strength of these kinematic signatures depends both on the relaxation time of the system and on the fractional mass of the 
second generation disk. We therefore conclude that SG populations formed in flattened configurations will give rise to two systematic trends: 
(1) Positive correlation between GC ellipticity and fraction of SG population (2) Positive correlation between GC relaxation time and ellipticity. 
Thereby GC ellipticities and rotation could be related to the formation of SG stars and their initial configuration.   
\end{abstract}


\section{Introduction}\label{Intro}

Observations over the last two decades have shown that globular clusters (GCs) are not monolithic objects composed { by a single stellar population}; 
they host multiple stellar {  populations} 
of stars.
The different  populations are characterized by
chemical inhomogeneities in light elements, like He, C, N, O, F, Na, Al, Mag and Si, and by
a strong anti-correlation between Na and O \citep{2004ARA&A..42..385G, 2012A&ARv..20...50G}.
 For convenience, we will call stars with an higher content of He, Na and Al and a lower content of C and O 
``second generation'' (SG) stars.
SG stars are thought to account for 
a large fraction of the cluster total mass, typically from $\sim$30\%-70\% of the total number of stars 
\citep{2008MNRAS.390..693D, 2009A&A...505..139C, 2009A&A...505..117C, 2010A&A...524A..44P, 2015MNRAS.453..357B}. 
{ \cite{2015MNRAS.453..357B} recently argued that the fraction of SG stars is remarkably uniform, $\sim70\%$,
apart from a few exceptions, and that the earlier reported smaller fractions might be considered as lower limits. }
The origin of {  SG stars} is still debated, but they are typically thought to arise from gas accreted into the GC core which 
then forms a new generation of stars at its central parts. The origin of the gas could be either internal or external, 
and can originate, for example, from material ejected by primordial, first generation (FG) asymptotic giant branch 
(AGB) stars \citep{2001ApJ...550L..65V, 2008MNRAS.391..825D}, {  interacting} massive FG binaries \citep{2009A&A...507L...1D,2013MNRAS.436.2398B},
{  very massive stars \citep{2014MNRAS.437L..21D}
or fast rotating massive stars \citep{2007A&A...464.1029D, 2013A&A...552A.121K}}. Material ejected at 
sufficiently low velocities (lower than the escape velocity from the GC) can be retained in the cluster, concentrate 
at its center and mix with the leftover pristine 
gas and then fragment to form new stars \citep[see][and references therein]{2012A&ARv..20...50G}. Different sources 
imply distinct He enrichment with respect to that of Na and a range of age differences between FG and SG stars, up to few hundred Myr in the case of the AGB polluter scenario { \citep[see e.g.][]{2010MNRAS.407..854D,2013A&A...552A.121K}.}
{  In order to explain the large SG fraction the initial cluster should have been initially 10-100 more massive \citep[see][and references therein]{2012A&ARv..20...50G}, this issue known as the ``mass budget problem''. \\}
{  The recently proposed early disc accretion  scenario  \citep{2013MNRAS.436.2398B}, where young low mass stars accrete { intra-cluster gas,
polluted by massive interacting binaries} through their { circumstellar} discs, without requiring a second star formation epoch,
solves the mass budget problem but, as any of the self-enrichement scenarios, fails to simultaneously 
explain the He and Na enrichment in most GCs \citep{2015MNRAS.449.3333B}.} 
{  In all the self-enrichment scenarios}, the initial distribution                                     
of the newly formed stars strongly depends on the gas configuration. 
Recent simulations \citep{2010ApJ...724L..99B, 2011MNRAS.412.2241B} showed that the dissipative 
accretion of AGB gas, that 
carries a small amount of rotation inherited from the parent FG stars, can lead to  
the formation of a second generation gaseous disk rather than a spherical distribution of gas. Fragmentation of 
such a disk would give rise to a stellar disk embedded in the spherical cluster. The interaction between these two 
stellar structures and the later cluster relaxation could significantly differ from that of a spherical cluster of a single 
stellar population and from that of a cluster with an embedded spherical cluster of SG stars. \\
In  \cite{MBP14} (hereafter Paper I) we began to study the evolution of such embedded SG disks, and focused on the test 
case of an $\omega$ Cen-like cluster. We found that the presence of the disk leaves kinematical and morphological signatures 
on the structure of the cluster. These features are potentially still detectable even after 12Gyr of evolution.  
The initially spherical components is slightly flattened while
the disk stars do not completely mix with the FG population. 
The velocity dispersions of the FG and the SG populations still differ after 12Gyr;  
the FG population shows a mild tangential anisotropy while the SG population shows significant radial anisotropy.

Here we extend our analysis of the evolution of SG disks to the more general case and consider
SG disks with a range of masses.
{  A { small} spread in number ratio between the first and second generation stars  has been recently observed
\citep[see e.g.][]{2015MNRAS.453..357B}. 
{ More importantly, } although the presence of young stellar 
disks in NSCs seems to be ubiquitous, the mass of their different stellar generations is not known \citep{2015AJ....149..170C}.}
{  Thus we} explore the effects of these disks on the cluster evolution, and the dependence of the cluster 
properties on the fractions of SG stars in the disk. We also compare these results to the case of SG stars formed 
in an embedded spherical sub-cluster rather than a disk configuration.  

We note that recently \cite{HBG15} have also studied SG formation scenarios and considered formation in SG disks. 
When relevant we compare our results with their findings.  

In Section \ref{sec:meth} we describe the methods used and the simulations run. The results of these simulations 
are shown in Section \ref{sec:resu}  their observable implications are and discussed in Section \ref{sec:comp}, followed by the summary
and conclusions in Section \ref{sec:disc}.
\begin{figure*}
\begin{center}
\includegraphics[width=0.9\textwidth,clip=true]{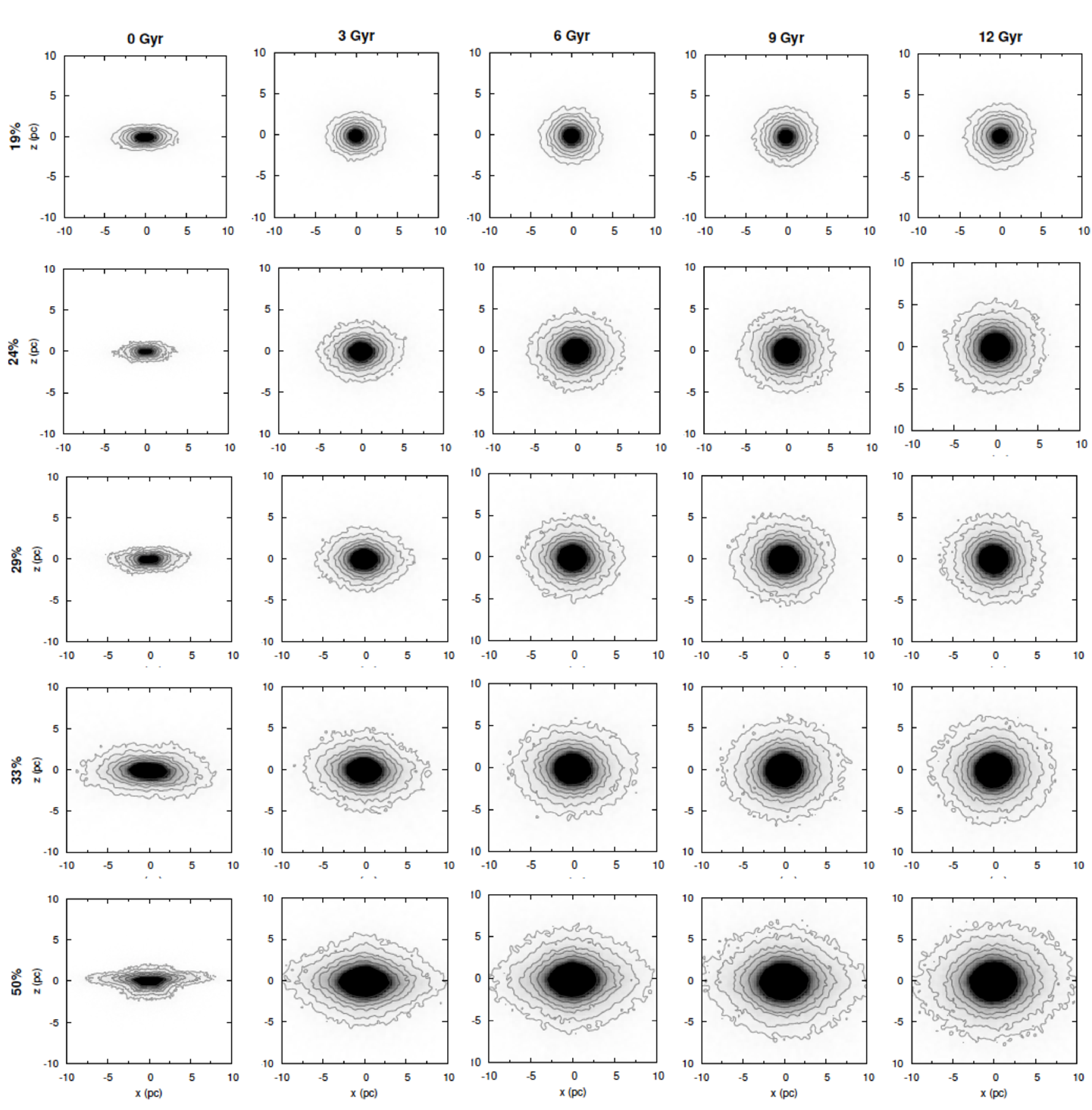}

\caption{The evolution of the  SG simulated disks with time. The isodensity contours are given at 0, 3, 6, 9 and 12 Gyr (going from the left to the right) and 
for the 19\%, 24\%, 29\%, 33\%, 50\% mass fraction disks (going from top to bottom). The density contour levels are
50, 100, 200, 300, 500 and 1000~$M_\odot/$pc$^3$.
The $z$ axis is parallel to
the $L_z$ component of the disk angular momentum. The disks clearly inflate with time, becoming almost
spherical after 12Gyr. Different systems have different relaxation times so their final configuration slightly depends also on this factor.}
\label{fig:disks_fin}
\end{center}
\end{figure*}
\begin{figure}
\begin{center}
\includegraphics[width=0.45\textwidth, trim=1cm 0.5cm 1cm 1cm,clip=true]{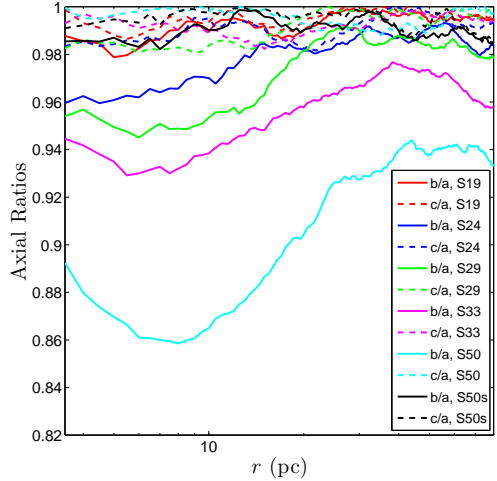}
\end{center}
\caption{Axial ratios of both first and second generation populations { after 12Gyr of evolution}.}
\label{fig:axialratios}
\end{figure}
\begin{figure*}

\begin{center}
\includegraphics[width=0.27\textwidth, trim=2.5cm 0cm 2.5cm 1cm,clip=true]{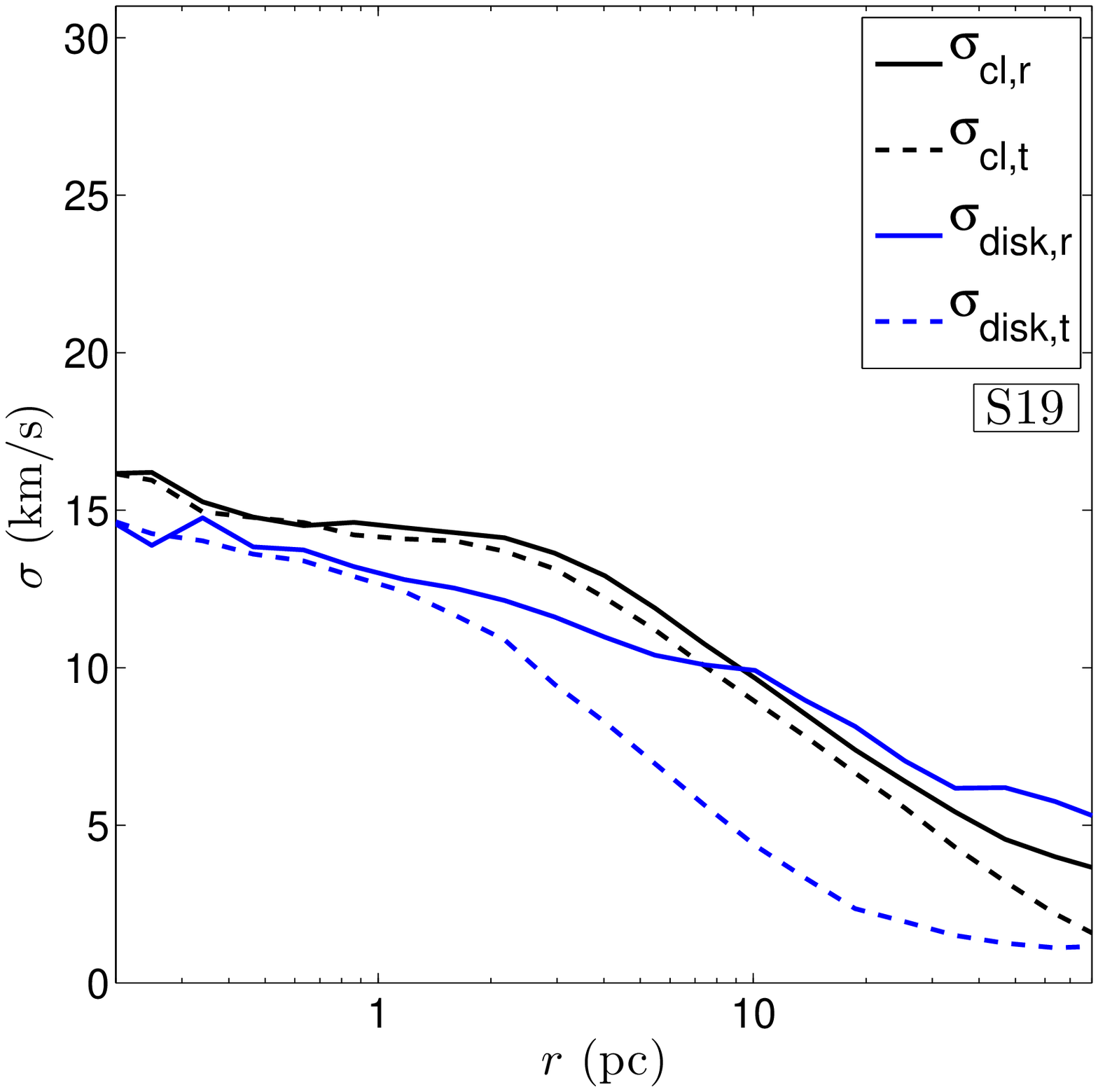}
\includegraphics[width=0.27\textwidth, trim=2.5cm 0cm 2.5cm 1cm,clip=true]{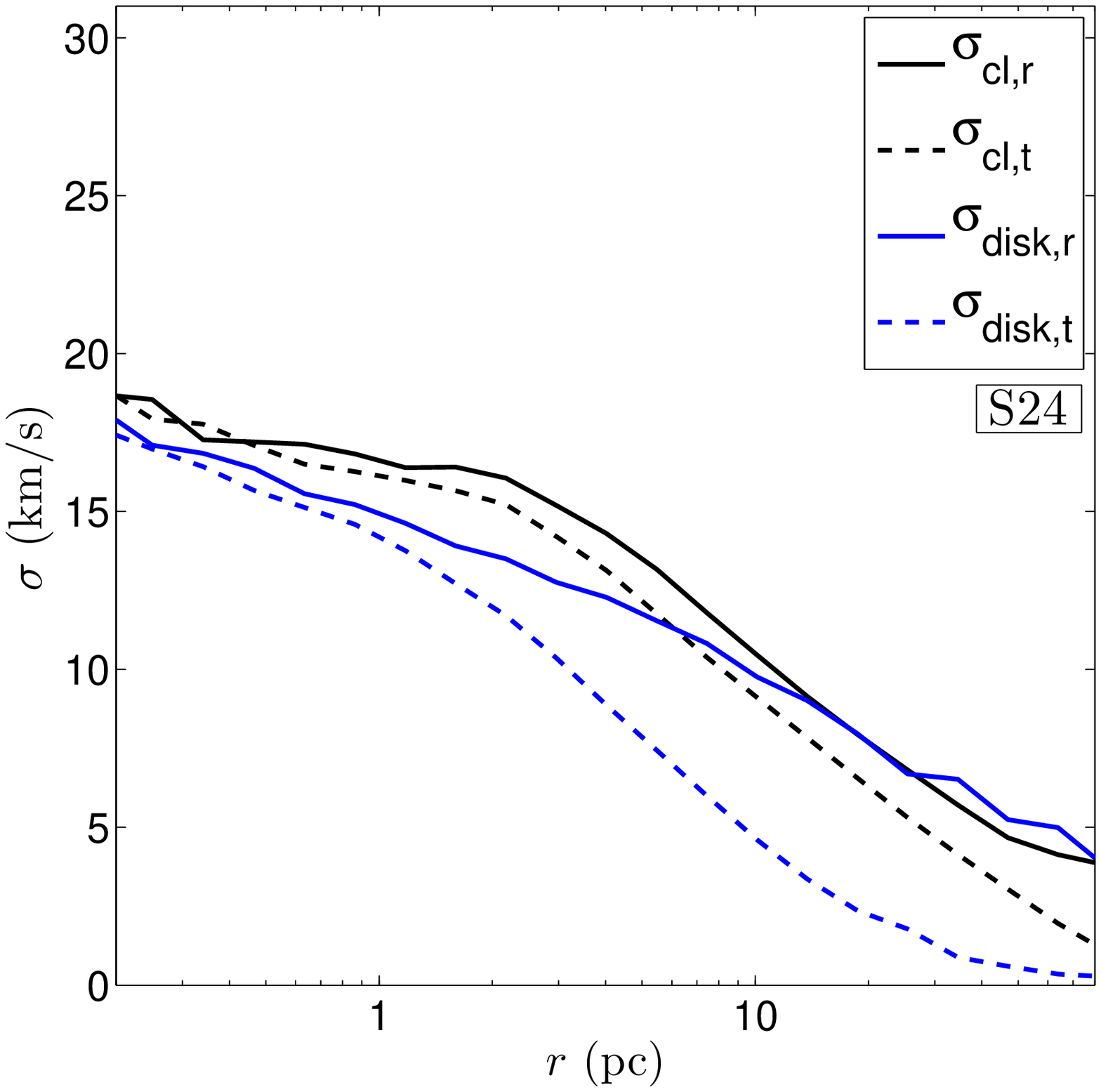}
\includegraphics[width=0.27\textwidth, trim=2.5cm 0cm 2.5cm 1cm,clip=true]{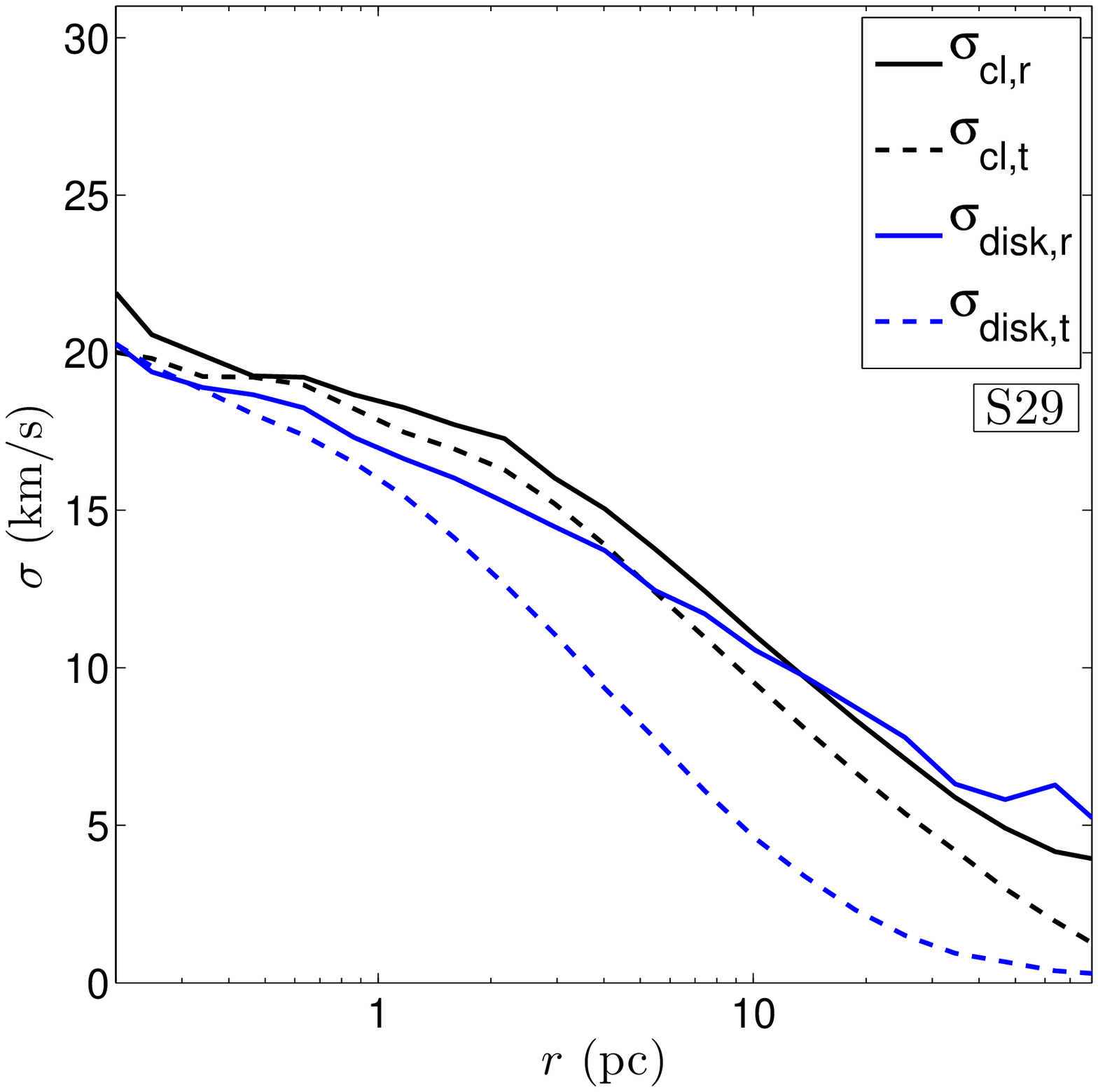}\\
\includegraphics[width=0.27\textwidth, trim=2.5cm 0cm 2.5cm 1cm,clip=true]{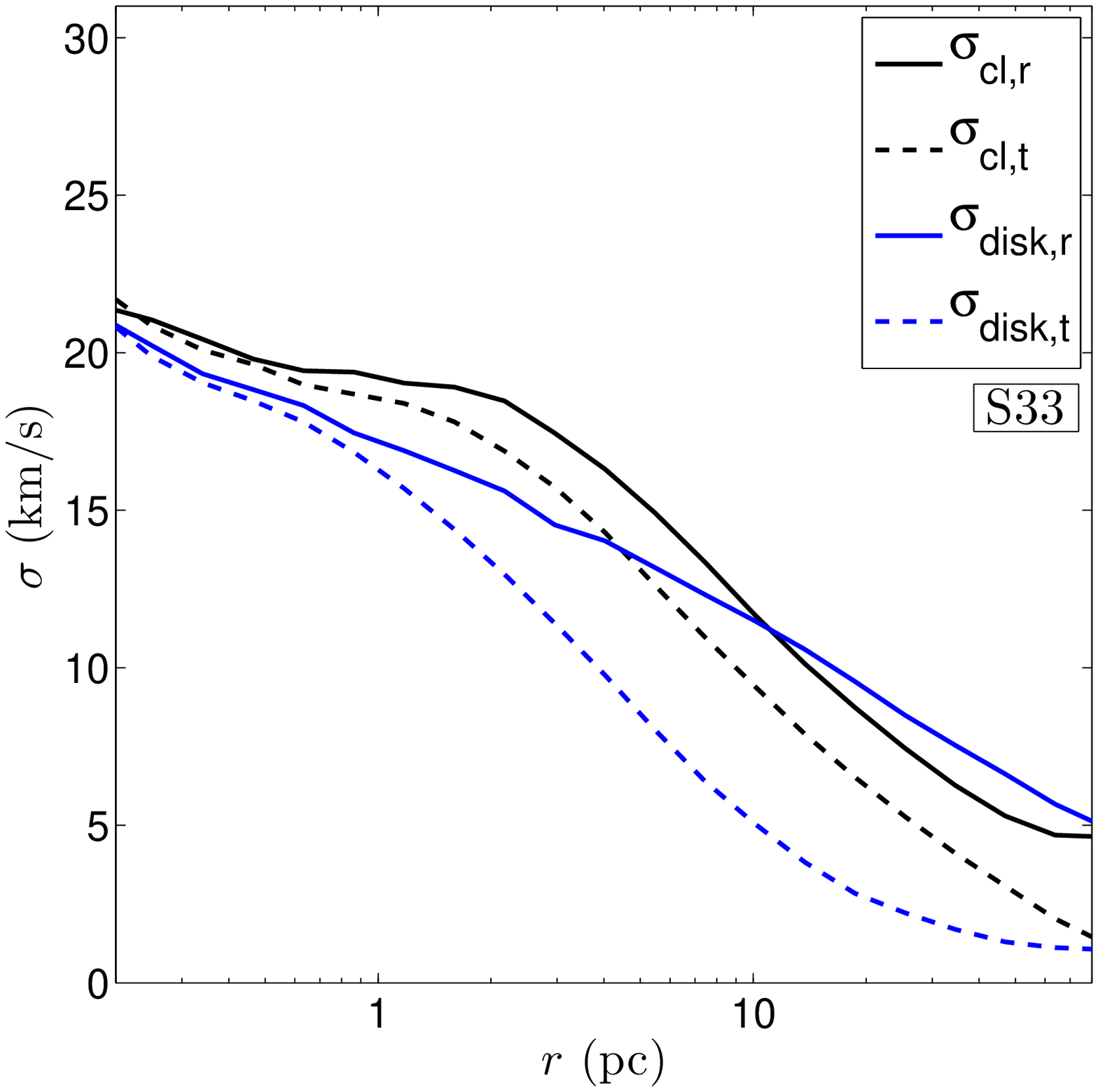}
\includegraphics[width=0.27\textwidth, trim=2.5cm 0cm 2.5cm 1cm,clip=true]{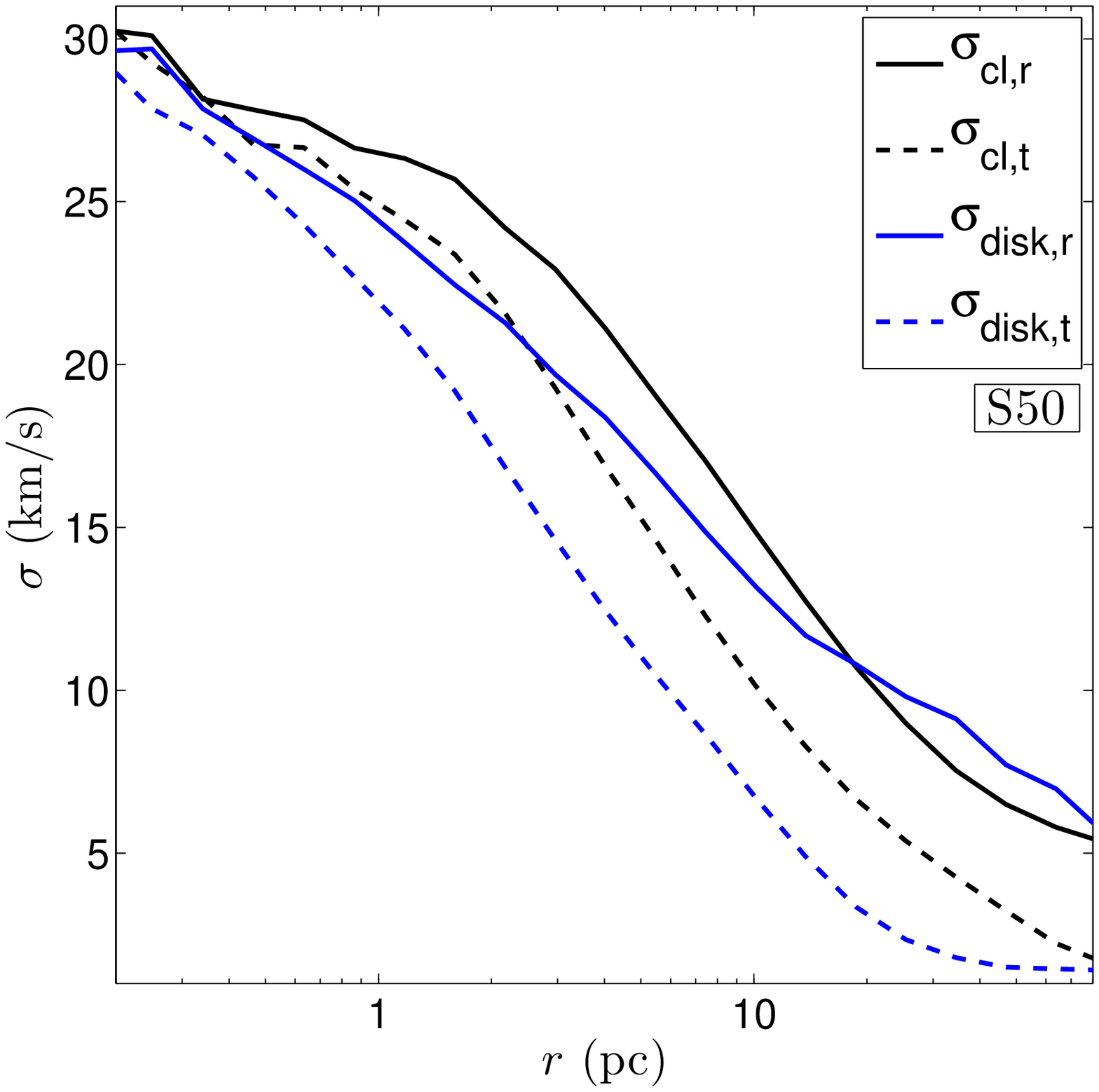}
\includegraphics[width=0.27\textwidth, trim=2.5cm 0cm 2.5cm 1cm,clip=true]{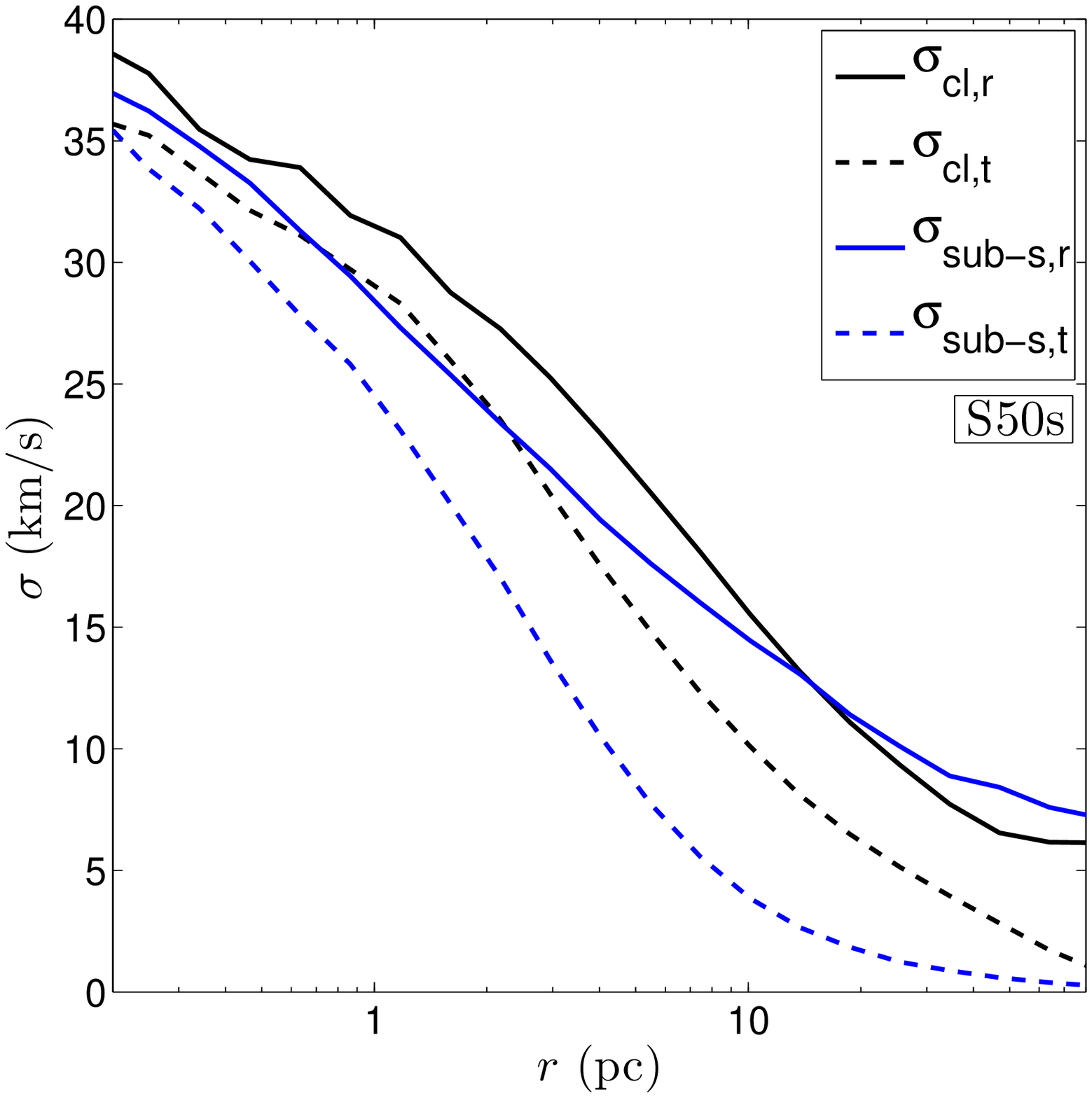}
\caption{Radial (solid line) and tangential (dashed line) velocity dispersions for the different simulated systems, { after 12Gyr of evolution}. The black lines are for the FG population (cluster)
and the blue ones are for the SG population (disk).} 
\label{fig:sigma}
\end{center}
\end{figure*}

\section{Methods}\label{sec:meth}
The evolution of massive second generation disks in GCs has been explored by means of high-precision N-body simulations.
In the following we describe the methods and tools used (similar to those used in Paper I), 
as well as the various initial conditions (summarized in Table \ref{tab1}).
\subsection{The N-Body Simulations}
We run several N-body simulations of stellar disks embedded in an initially spherical cluster, 
using $\phi$GPU \citep{Ber11}, 
a direct N-body code running on graphic processing units (GPUs). 
The code has been developed based on $\phi$GRAPE, a previous code optimized to use GRAPE6 cards
as hardware accelerators.
$\phi$GPU code is fully parallelized using the MPI and CUDA C, the language developed to program NVidia GPUs.

We run our simulations both on the cluster Tamnun at the Technion
and on the cluster Cytera in Cyprus.
Tamnun is equipped with 4 nodes with one Nvidia Tesla M2090 each and other four nodes with 
three Tesla K20m. Cytera is an HPC cluster with 18 GPU nodes with dual NVidia M2070 GPUs.
In order to smooth close encounters between particles we used a softening length of
$10^{-3}$~pc. As shown in Paper I even a softening length as big as $5\times 10^{-2}$~pc
does not affect the results significantly.
The relative energy variations at the end of the simulations are smaller than  $\Delta E/E=0.01$.
\begin{table}
\begin{center}
\caption{Number of particles in the FG and SG populations  and relative initial half-mass relaxation times. }
\label{tab1}
\begin{tabular}{cccc}
\hline
\hline
ID & $N_{FG}$& $N_{SG}$ & $t_{rh,0}$ (Gyr)\\
\hline
S19 & 50000 & 11175 & 7.0\\
S24 & 50000 & 16121 & 7.0\\
S29 & 50000 & 19193 & 8.2\\
S33 & 50000 & 25069 & 8.5\\
S50 & 50000 & 45865 & 8.3\\
S50s & 50000 & 50000 & 8.6 \\
\hline
\end{tabular}
\end{center}
\end{table}
\begin{figure*}
\begin{center}
\includegraphics[width=0.4\textwidth, trim=8cm 0cm 7cm 0.5cm,clip=true]{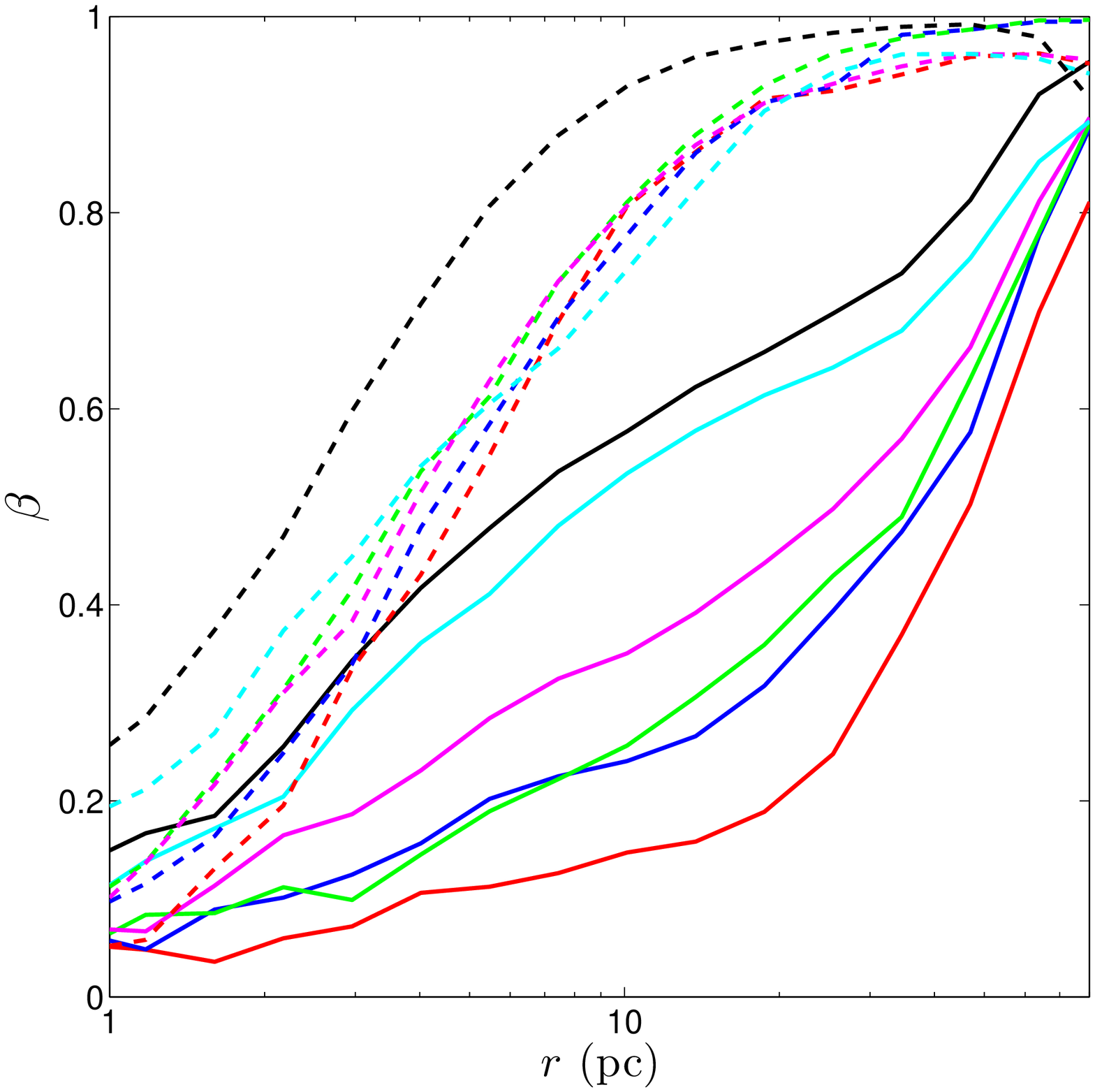}
\includegraphics[width=0.4\textwidth, trim=8cm 0cm 7cm 0.5cm,clip=true]{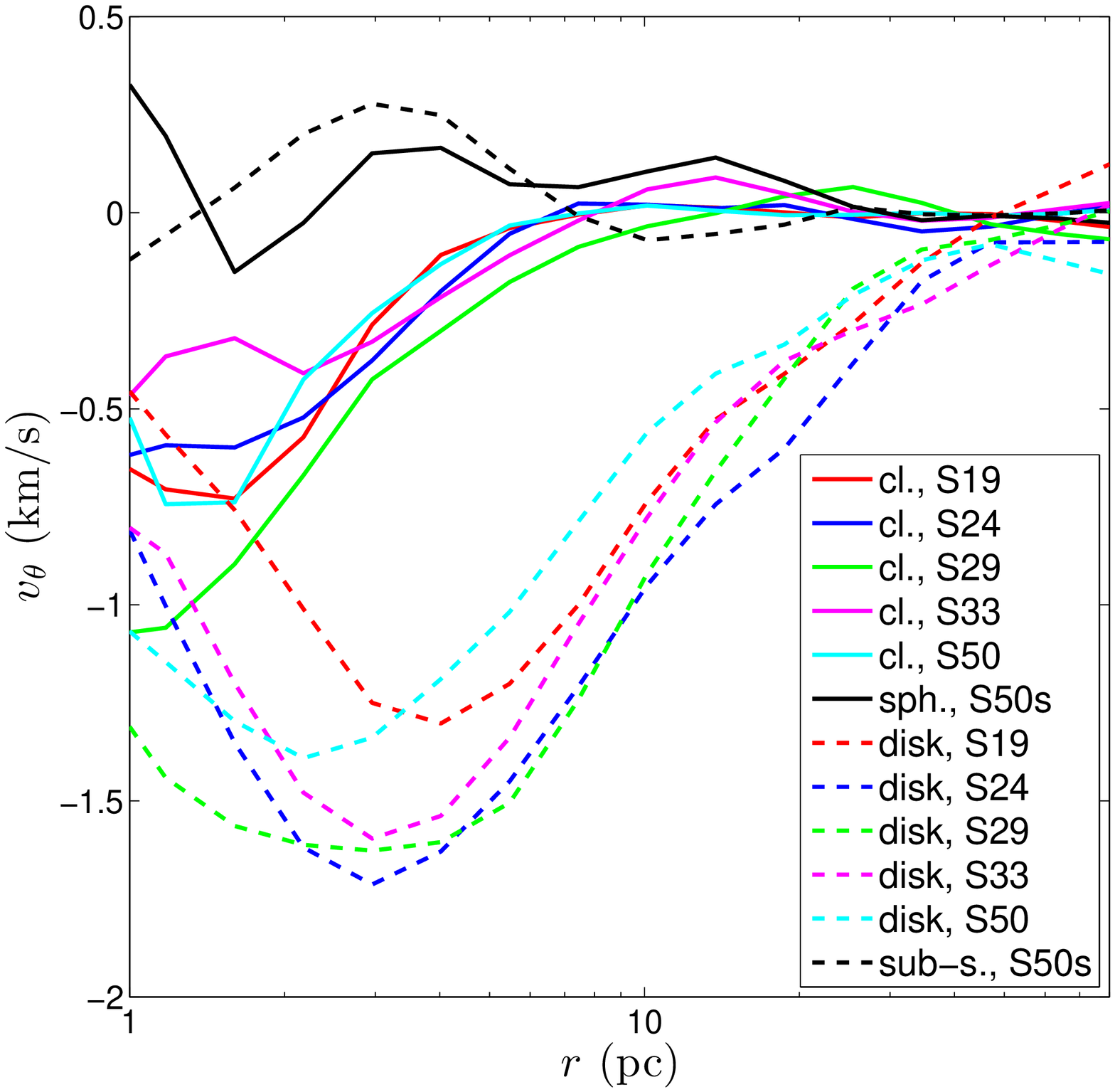}
\caption{The final velocity anisotropy parameter (left panel) and the azimuthal velocity (right panel)
for all the systems. The solid lines are for
the initially spherical component, the dashed lines are instead for the flattened or spherical second generation population. }
\label{fig:beta}
\end{center}
\end{figure*}
\begin{figure}
\begin{center}
\includegraphics[width=0.45\textwidth, trim=15cm 0.cm 15cm 1cm,clip=true]{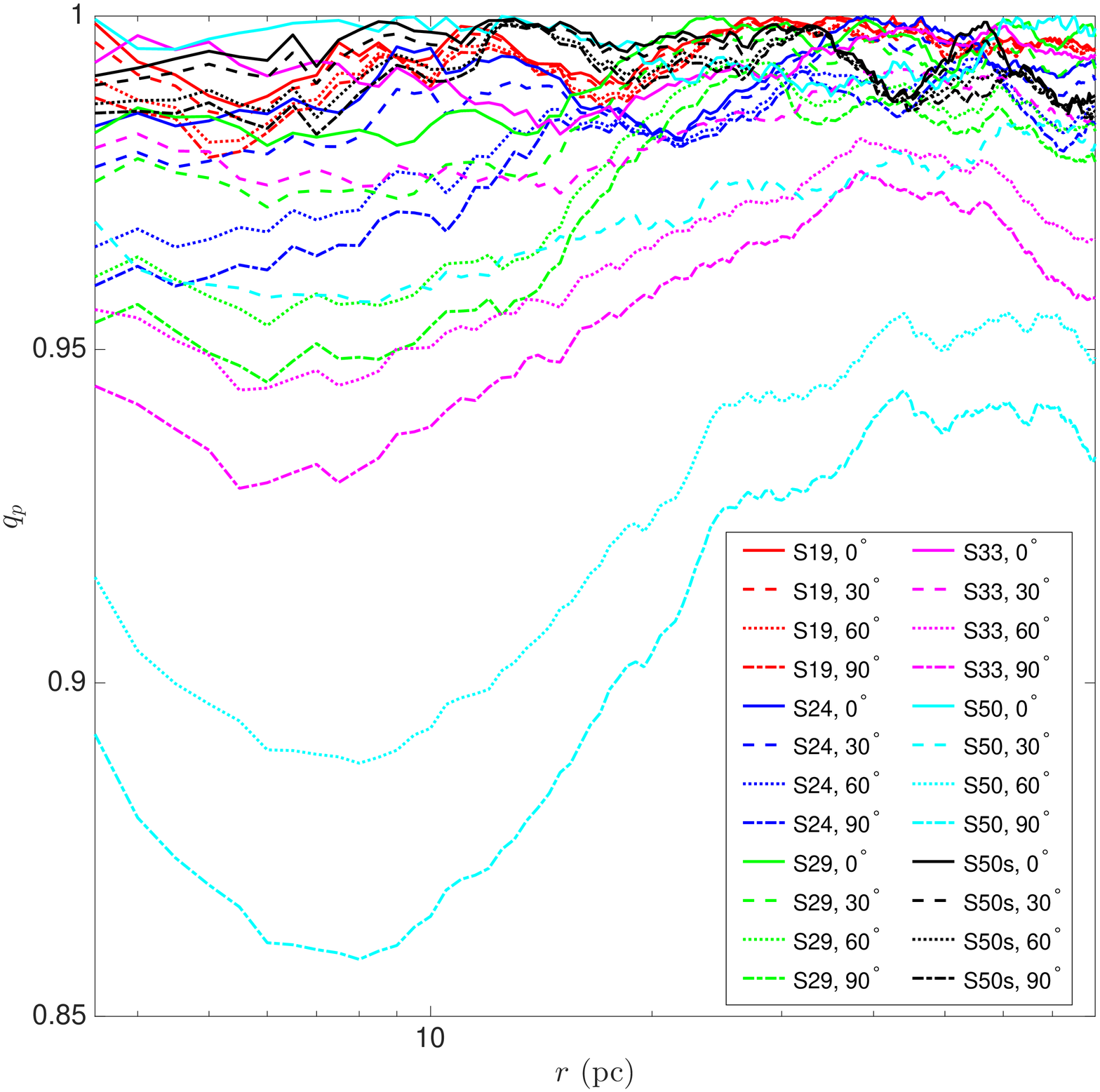}
\end{center}
\caption{Projected axial ratios of the simulated systems at the end of the runs.}
\label{fig:axialp}
\end{figure}

\subsection{Initial Conditions}
The initial spherical, FG, component of the cluster is 
a single mass 
King model, similar to the best fit model of the density profile of $\omega$ Cen \citep{1987A&A...184..144M},
with $W_0=6$, core radius $r_{c}=4.4$~pc and tidal radius $r_{t}=80$~pc.
This system is sampled using $N=50000$ particles, whose mass is $20~M_{\odot}$, accounting for a total mass of
$M=10^{6}~M_{\odot}$.

We embed each of the SG disks we consider in the initially spherical FG stellar population. 
{   We assume that the cluster has already lost a large fraction of FG stars and we focus on its the long term evolution.}
The disk is allowed to relax inside the cluster until it reaches a quasi-stable state, which is then considered to be the initial configuration of the system.
During the pre-initial stage a fraction of the SG stars in the disk are expelled from the system, and are removed from the N-body simulation, 
leaving behind a disk whose mass corresponds to 19\%, 24\%, 29\%, 33\% and 50\% of the total  mass of the system (first+second generation populations), 
in the different models considered (see Table \ref{tab1} for more details
 and for the simulations IDs used throughout the text).
{  The profile of the FG component, after the initial relaxation phase, is still well described by the initial parameters. 
At the same time the disk inflates achieving a flattening q=0.4; 
90\% of its total mass is inside the core radius of the FG spherical cluster. }
The initial half-mass relaxation time, $t_{rh}$\footnote{$t_{rh}$ has been evaluated using the definition given by \citep{1987degc.book.....S}, 
estimating the number density and the root mean square velocity. {  This expression of the relaxation time is in principle only valid for spherical and isotropic systems. For a system with coherent velocities (such as a disk), the relaxation time would be shorter \citep[see e.g.][]{2014PhDT.........4H}. The half-mass relaxation times listed in Table  \ref{tab1}   may therefore be considered as upper limits for systems with an embedded disk.}}, of each simulated system is also given in Table \ref{tab1}.
{   The FG spherical component needs to rotate to form a SG disk \citep{2010ApJ...724L..99B}. However, in our case, the  initial disks have a maximum rotational velocity of $\sim 7$km/s. Since this value  corresponds to a peak rotational velocity smaller than ~0.5km/s for the FG population (see Figure 3 in \citealt{2010ApJ...724L..99B}), we neglect this initial effect.}
 N-body simulation of GC systems are computationally expensive, and we therefore use a smaller number of particles to represent the FG cluster and SG disk stars, 
 as to allow for faster computation. The mass of each N-body particle is therefore larger  than the typical average stellar mass in GCs. 
 Since this choice affects the dynamical time-scales, we rescale the simulation time to the relaxation time of the corresponding real system, 
 using the procedure described in Paper I, assuming an average stellar
mass of $\langle m_{*}\rangle=0.5~M_{\odot}$ for the stars in a realistic GC. 
Each system is evolved for 12Gyr of rescaled simulation time, comparable to the age of GCs.

The dynamics of stars born in a disk can give rise to different signatures observable in the cluster structure, 
compared with the effects of SG stars born in a spherical sub-cluster.
In order to compare these two different initial configurations, we run an additional simulation where the SG stars are distributed inside the core radius
of the FG population with a King profile of the same parameters as used for the FG stars. 
The latter case is simulated only for a SG population comprising 
50\% of the total mass of the cluster (including both FG and SG stars).

The clusters are simulated in isolation; tidal effects due to the galaxy significantly affect 
only the most external regions of the cluster, and are neglected here\footnote{  In order to check that the effects of an  external potential can be neglected
we run an additional simulation with the cluster orbiting a point-mass galaxy with the same mass of the Milky Way within 8.5~kpc, the cluster being on a circular orbit at this radius. The final system (we ran only the case with a mass ratio of 33\% between the two populations) shows the same final features of its counterpart evolved in isolation. }.
{  We did not take into account stellar evolution. Generally, mass loss would slow down relaxation, and stronger kinematic signatures 
should be left. We did not consider a mass spectrum for the stars, which could also affect the relaxation time, however, massive stars 
are short lived (and stellar black holes are rare) and most of the cluster evolution would be dominated by the  dynamics of low mass stars, for which the range of masses is modest.
}

\section{Results}\label{sec:resu}
We analyzed the morphological and kinematical properties of the system after 12Gyr in order to find how the initial presence of a disk modifies
the structure of the whole cluster, leaving behind long term observable signatures. 

\subsection{Morphological properties}\label{sec:morph} 
Figure \ref{fig:disks_fin} shows the isodensity contours of each stellar disk at the beginning of the simulation and after 3, 6, 9 
and 12Gyr. 
{  As already found in Paper I and by \cite{Ve13}}, the second generation stars always remain confined within the central 10-20pc without completely mixing with the FG population.
The 19, 24 and 29\% disks, at the end of the simulation, are almost spherical while the 33\% and the 50\% disks are
still significantly flattened. We also note that the 50\% disk is unstable and forms a central bulge that could lead to a faster 
angular momentum exchange with the FG population through collective effects rather than two-body relaxation {  (see also Paper I)}.\\
Figure \ref{fig:axialratios} shows the axial ratios of the whole system at the end of each simulation. 
The 19\% disk is slightly prolate, with all axial ratios $\sim 1$, while the other systems are oblate. 
The $c/a$ axial ratio is smaller for larger disks,
the $50$\% disk is almost perfectly oblate while the clusters with lower mass disks are more triaxial. 
The cluster is flattened at any radius, but the effect is larger in the central 10-20pc, where most of the 
disk stars reside. In contrast, the spherical SG system (model S50s) and its  host clusters system do not show any significant flattening, as expected. 

\subsection{Kinematic properties}\label{sec:kin}
Figure \ref{fig:sigma} shows the final tangential and radial velocity dispersions
of all the initially spherical components and their comparison to embedded disk cases.
Both the disk and the cluster (FG) stars are centrally isotropic. 

In every case the disk stars show a large radial anisotropy that increases with the distance from the center  
(see left panel of Figure \ref{fig:beta}).
This is true also for the FG population that, nevertheless, shows a milder radial anisotropy. 
The velocity dispersion of 
the SG population remains lower than that of the FG population
even after 12 Gyr, {  as also found for the low mass  disk simulated in Paper I}. 
This difference is more apparent in the intermediate/external regions of the system.
The level of radial anisotropy of the disk stars does not show a clear dependence 
on the initial disk mass. In contrast, as also seen in  Figure \ref{fig:sigma}, the FG population shows 
a distinct dependence of its radial anisotropy on the initial disk mass.

The S50s case is highly radially anisotropic  (see Figures \ref{fig:sigma} and \ref{fig:beta}). 
Both the first and second generation populations are more radially anisotropic than
the systems hosting an initially flattened second stellar generation. 
This induced radial anisotropy has been similarly found by
\cite{HBG15}, who attributed it to the initially more centrally concentrated configuration of the SG stars.
In addition, we find that the higher radial anisotropy is correlated with higher mass ratio between the SG and FG stellar populations.
{  This correlation is due to the higher density of the disk and shorter relaxation time in systems with a more massive SG in respect of the FG.}

The right panel of Figure \ref{fig:beta} shows the azimuthal
velocity $v_{\theta}$ of the FG and SG populations as a function of the radius. 
In all the cases the FG populations show a mild central rotation with velocity between 0.5 and 1 km/s.
The SG populations, regardless of the mass of the disk, rotate with $v_{\theta}$ between 
0.5 and $2$km/s 
(within the central $\sim20$pc). The rotation initially increases, up to ~3 pc, and then decreases with the distance
from the center of the cluster. 
As expected \citep[see also][]{HBG15}, the spherical case does not show any significant rotation.

\subsection{Projected axial ratios}

In order to directly compare the morphological properties
of the simulated systems to those of their observed counterparts 
we evaluated the projected axial ratio $q_p$ of each system
as \citep{2008A&A...487...75C}
\begin{equation}\label{eq:proq}
q_p^2=\frac{q_1^2q_2^2}{q_2^2\cos^2(i)+q_1^2sin^2(i)}
\end{equation}

where $q_1$ and $q_2$ are the 3D major and minor axial ratios and $i$ 
is the inclination of the line of sight in respect to the $z$ axis.
We evaluated $q_p$ for all the systems after 12Gyr of evolution (Figure \ref{fig:axialp}). 
We adopted four different inclinations: $0^\circ$, $30^\circ$, $60^\circ$ and $90^\circ$. We find that the projected 
flattening grows with the disk mass. Note, however, that the observed flattening strongly 
depends on the inclination angle in respect to the line of sight; in particular, even very flattened clusters could appear spherical if seen face-on. 
Inclinations {  of $60^\circ$ and $90^\circ$ (edge-on) } amplify the flattening compared with the other cases.

As already found for the spatial axial ratios, the projected axial ratios of the cluster hosting
a spherically distributed SG population are of order of unity and very similar, regardless of the adopted viewing angle. 

\begin{figure*}
\begin{center}
\includegraphics[width=0.45\textwidth, trim=18cm 0cm 19cm 0cm,clip=true]{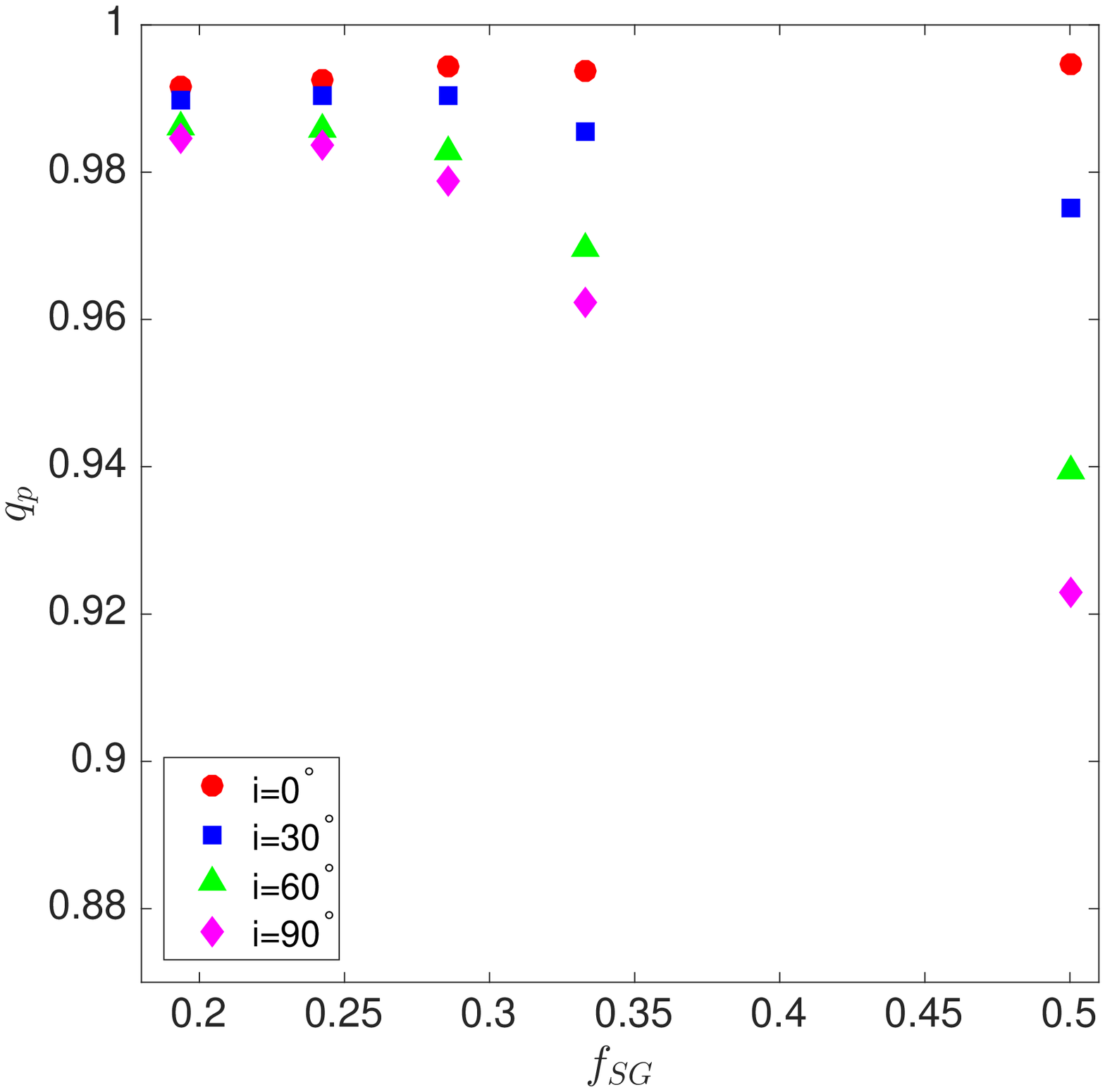}
\includegraphics[width=0.45\textwidth, trim=18cm 0cm 19cm 0cm,clip=true]{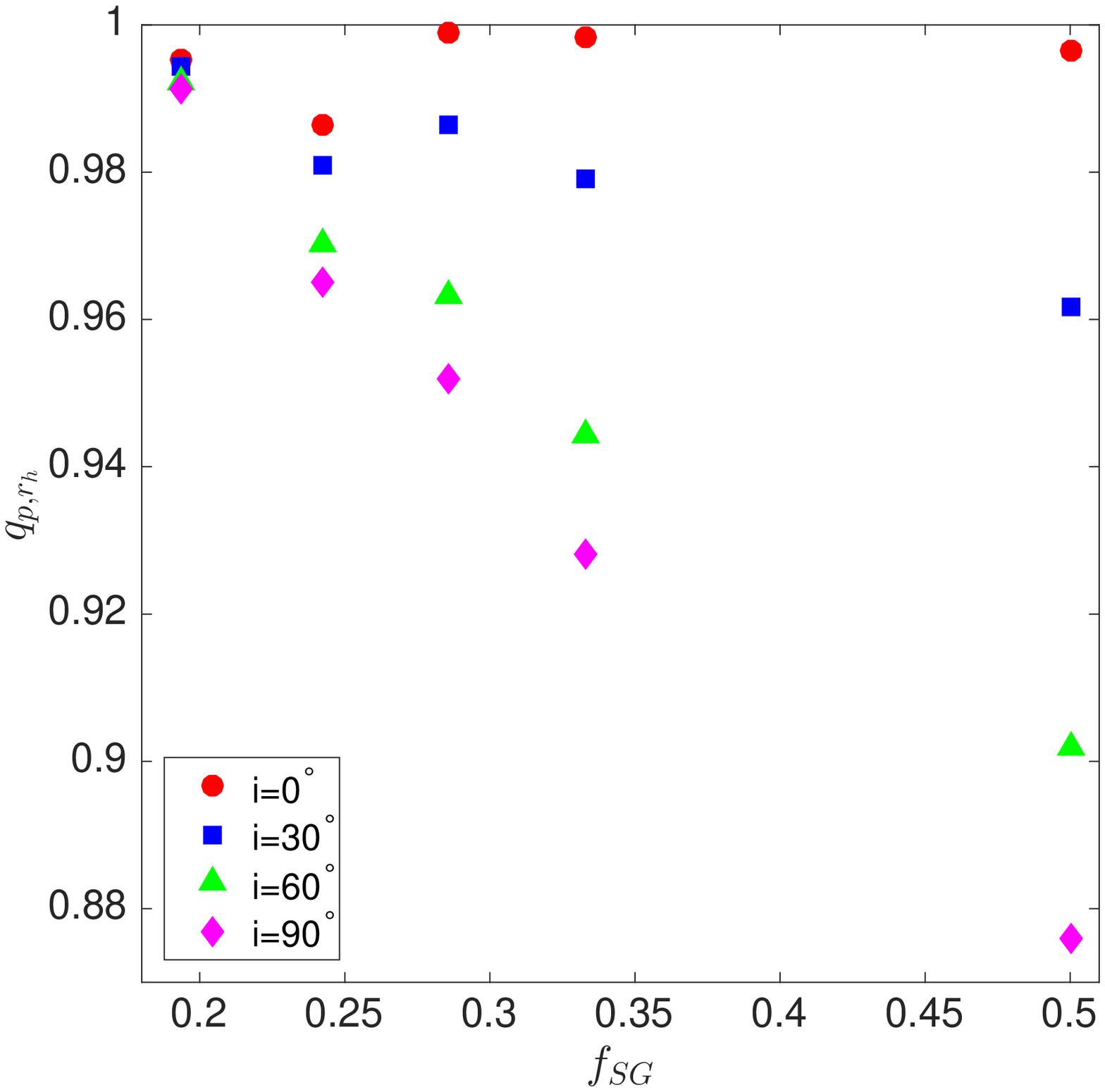}
\end{center}
\caption{Average  (left panel) and half mass radius (right panel) projected axial ratios 
of the simulated and observed GCs plotted against the SG mass fraction in respect to the total mass of 
the system. }\label{fig:ecc_frac}
\end{figure*}
\section{Observable implications}\label{sec:comp}
In the following we consider the observable signatures of second generation populations and their initial configurations on the current structures and 
properties of Galactic GCs. In particular we focus on specific trends expected as a function of the fraction 
of SG stars and the flattened vs. spherical configurations.
\subsection{Flattening}
Typical GCs are not spherical but are slightly ellipsoidal 
\citep[see e.g.][]{1983A&A...125..359G,1987ApJ...317..246W,2005MNRAS.360..631M,2010ApJ...721.1790C}. 
Several origins for the observed ellipticity has been suggested, including primordial internal rotation, 
external tides, and/or pressure anisotropy \citep{2006MNRAS.373..677F, 2008AJ....135.1731V, 2012A&A...540A..94V, 2013MmSAI..84..183B}. 
Mergers of globular clusters, that have been proposed as a possible explanation for
the presence of multiple stellar generations \citep{1996ApJ...471L..31V, 2010ApJ...722L...1C, 2013MNRAS.435..809A},
could also lead to significant flattening \citep{1991Ap&SS.185...63M}.
However, the absence of binary Galactic GCs \citep[see e.g.][]{1972MNRAS.160..249I,1991Ap&SS.183..129S,2008AJ....135.1731V} suggests that this process does not play an important 
role in the evolution of GCs in our Galaxy.
\cite{1991EM&P...54..203A} found that the external regions become more spherical with time as a consequence of gravothermal contraction 
and tidal stripping. At the same time, the tidal field can stretch the clusters, making them more elongate.
Moreover, \cite{1996AN....317..333S} found that moderate bright 
Galactic GCs are rounder than 
fainter and brighter clusters. 
The existence of a correlation between age and flattening 
\citep[][and reference therein]{1991Ap&SS.185...63M}
or between flattening and the presence of bluer horizontal branches  in clusters are still debated 
\citep{1983ApJ...272..245N, 2008AJ....135.1731V}.  \\

As also shown in Paper I, in our simulated clusters, the flattening is 
caused by the angular momentum exchange between the second and first generation stars.
When both the first and second generation are spherically symmetric, the system always keeps its initial morphology and does not rotate
(see Sections \ref{sec:morph} and \ref{subs:rot}). Note that since it does show 
significant radial anisotropy, the velocity anisotropies are not linked with the flattening (Sections \ref{sec:kin} and \ref{subs:veld}).

Since the central density and the angular momentum increase
with SG disk mass, both the total amount of the angular momentum transferred and the efficiency of the angular 
momentum exchange between the FG and SG populations 
are expected to be larger for more massive SG disks, as indeed observed in our simulations.

The 3D flattening is always larger than the projected one (see Figures \ref{fig:axialratios} and \ref{fig:axialp}),
thus observations can only provide a lower limit for this quantity.
However, statistical analysis and future observations, e.g potentially provided by the GAIA mission, could give more detailed information
on the spatial structure of GCs, allowing a quantitative comparison between the theoretical expectations and the observations. 

The projected axial ratios (see Equation (\ref{eq:proq})) of the simulated systems, averaged over the radial extent  and evaluated at the half mass radius, are 
shown in Figure \ref{fig:ecc_frac} as function of the SG mass fraction of the total mass 
of each system and for four different projection angles ($0^\circ$,$30^\circ$, $60^\circ$, $90^\circ$).
If seen face-on ($0^\circ$) all the clusters look almost spherical. Different inclinations produce different values
of the flattening, and this could be connected to the large spread in $q_p$ for the observed Galactic GCs (see Figure \ref{fig:ecc_frac}).
The trend of ellipticity correlation with SG mass fraction can be clearly seen.
\subsection{Velocity anisotropies: a tracer for the mass of second generation stellar population}\label{subs:veld}
Both the cluster hosting a spherical SG population and the clusters whose SG stellar populations have initially flattened structures show 
radially anisotropic profiles. Therefore, this feature cannot be used as a tracer for the initial SG configuration 
\citep[see also][]{HBG15}, however it can be used as an indicator of the total mass of the SG population.
The simulated clusters have similar relaxation times and their radial anisotropy grows with time. After 
12Gyr of evolution, the larger the second generation mass (independent of the initial configuration),
the larger is the radial anisotropy of the FG population. The SG population is more radially
anisotropic than the FG one. 

One difference between the spherical SG case and the flattened SG populations is 
that even when the mass of the spherical SG model is similar to that of the most massive simulated disk, 
its velocity dispersion is more radially biased than that found for the S50 case. 
This is { possibly} caused by  the faster relaxation in the case of an initially flattened distribution for the  SG population.
\footnote{ If a disk-like structure relaxes faster than a spherical one \citep[see e.g.][]{2014PhDT.........4H}, the radial
anisotropy due to the initially more concentrated configuration is expected to be erased earlier in time 
for an initially flattened SG population.}

\subsection{Rotation: the disk signature}\label{subs:rot}
A bulk average rotational velocity has been already observed in few Galactic GCs. Recently \cite{2013ApJ...772...67B} and \cite{2014A&A...567A..69K}
found that the rotation plays an important role in the shaping the structure of $\omega$ Cen, M15, 47 Tuc and NGC 4372
and it gives rise to the flattening of these clusters. Furthermore, \cite{2014ApJ...787L..26F} found a tight correlation between rotation and flattening for
the 11 Milky Way GCs in their sample.\\
The FG population in our simulation rotates with $v_{\theta}$ smaller than 1km/s.
The SG population have a maximum $v_{\theta}$ between 1km/s and 2km/s, and all the
systems reach the maximum velocity at $\sim$3pc from the center of the cluster, regardless of the mass of the disk.
The rotation is apparent even beyond 10pc -  the radius containing most of the second generation stars. 
{  As a consequence of how the initial conditions are built the} spherical S50s system does not show any significant rotation in any of the components.
Thus, in  the AGB model or in any model where the SG stars form in a disk, the  rotation can be potentially explained as
the result of angular momentum exchange between an 
initially flattened and slowly rotating SG stellar population and an initially spherical FG population. 

\subsection{Core collapse}\label{subs:cc}
The strength of the signatures left by the SG disk depends both on the mass of the disk and on the relaxation time of the system, 
but all the signatures are evident even after 12Gyr in any of the simulated cases.
All the systems studied have relaxation times shorter than a Hubble time and the presence of disks induce 
a more rapid core collapse of the cluster; much faster than that expected from non-centrally concentrated SG population.
More massive disks lead to faster core collapse, and we therefore expect clusters with a larger second generation 
fraction to be more dynamically evolved than their counterpart similar systems with smaller second generation fractions.

\section{Summary and Conclusions}\label{sec:disc}
We have explored the long term evolution of massive second generation disks embedded in an
initially spherical cluster composed of primordial, first generation stars. We have also compared these results with the 
case of SG stars in a spherical configuration. 
The initial existence of a SG disk gives rise to kinematic and structural signatures imprinted in the cluster properties. 
These signatures are potentially observable even today after 12Gyr of evolution, 
and the strength of these signatures is correlated with the SG population mass fraction. 
These signatures become weaker and less observable due to relaxation processes, and should therefore be more pronounced in clusters with 
longer relaxation times; nevertheless, observable signatures can still linger even after more than a relaxation time.

Our main findings can be summarized as follows
\begin{itemize}
\item Similarly to what found in Paper I {  and in \cite{Ve13}}, second generation stars remain  confined in the central 10-20pc and never completely mix with 
FG population. This is still true even after more than one (initial) relaxation time of the system.
\item The evolution of the disk inside the cluster leads to a significant flattening of the system (see Paper I as well). The amount of flattening increases 
with the mass of the SG disk.
\item The least massive system becomes slightly prolate while the other clusters are significantly oblate,
especially in their central regions. However, the more massive systems are oblate up to their tidal radius.
Therefore, future observations of the external, less crowded, regions of GCs may provide us with important clues regarding 
the early star formation history in these clusters.
\item The most massive system is also the most flattened one (see Figure \ref{fig:axialratios}). Its $c/a$ ratio is 
consistent with that of the most flattened GCs observed in the Milky Way \citep{1996AJ....112.1487H}.
\item The projected axial ratios strongly depend on the viewing angle and increase with the mass of the SG disk.
\item The spherical component,
as well as the whole system, becomes radially anisotropic as the disk inflates {  \citep[see also][]{HBG15}}. 
In particular, the more massive the disk is, the more radially anisotropic the cluster becomes.
\item As already observed in Paper I, the FG population shows a larger velocity dispersion 
compared with the SG population. 
\item This is true also when the SG has an initially spherical configuration. The anisotropy is a tracer of the mass of the second generations
and not of their initial configuration.
\item {  Confirming what found by \cite{HBG15}}, the rotational features of the two populations are remarkably distinct. While the FG population does not rotate, 
at least outside the central few parsecs, 
the SG population does show rotational features, with a peak in the azimuthal velocity around 3pc from the cluster center. 
The more massive the SG population the larger is the overall cluster ellipticity.
Future observations of differential rotation between first and second generation stars could
then tell us about the initial configuration of the second stellar generation, 
shedding light on its formation mechanism.
\item The presence of a central denser population speeds up the core collapse.

In conclusion, the contemporary observation of flattening, lower velocity dispersion for the SG stars, radial anisotropy and SG
rotation  could point to an initially disk-like configuration of the younger, { light-element} enriched populations. 
The strength of  these signatures could provide information on the dynamical age of the clusters and also on the fraction of SG stars present in the system.
\end{itemize}
\acknowledgments{
The authors wish to thank the referee for helpful comments
and suggestions.
This research was supported by the I-CORE
Program of the Planning and Budgeting Committee and The Israel
Science Foundation grant 1829/12. We acknowledge the Cyprus Institute under the LinkSCEEM/Cy-Tera
project. }
\bibliographystyle{aa}
\bibliography{disks}

\end{document}